\documentclass[graybox]{svmult}

\usepackage{type1cm}

\usepackage{makeidx}         
\usepackage{graphicx}       
                            
\usepackage{multicol}       
\usepackage[bottom]{footmisc}

\usepackage{amsmath}
\usepackage{bbm}
\usepackage{dsfont}
\usepackage{booktabs}

\usepackage{newtxtext}      
\usepackage{newtxmath}       

\usepackage{url}
\usepackage{amssymb}

\makeindex

\begin{document}

\title*{A Regulated Market Under Sanctions. \\ On Tail Dependence Between Oil, Gold, and Tehran Stock Exchange Index}
\titlerunning{A Regulated Market Under Sanctions}
\author{Abootaleb Shirvani, Dimitri Volchenkov}
\authorrunning{Abootaleb Shirvani, Dimitri Volchenkov} 
\institute{Department of Mathematics and Statistics,
	1108 Memorial Circle, Lubbock, TX 79409, USA, \email{abootaleb.shirvani@ttu.edu,dimitri.volchenkov@ttu.edu.}}

\maketitle


\abstract{
	We demonstrate that the tail dependence should always
	 be taken into account as a proxy for systematic risk of loss for investments.  
	We provide the clear statistical evidence of that 
the structure of investment portfolios on 
a regulated market 
should be adjusted to the 
price of gold. 
Our finding suggests that the active bartering 
of oil for goods would prevent collapsing
the national market facing the international sanctions. 
 }

\begin{flushleft}
		\end{flushleft}

\section{Introduction}

Since 1979, the UN Security Council and the United States regularly passed a number of resolutions imposing economic sanctions on Iran regarding  supporting for Iran's nuclear activities \cite{Guzman:2013}. Over the years, sanctions have taken a serious toll on Iran's economy and people, 
as well as resulted in the increasing government control over the forces of supply and demand  and prices in Iran.

Oil price changes 
have a significant effect on economy since 
oil prices directly affect the prices of goods
and services made with petroleum products.
Increases in oil prices can depress the supply of other goods, 
increasing inflation and reducing economic growth \cite{Blanchard:2007}.
Being an energy superpower, 
 Iran has an estimated 158 bln barrels of proven oil reserves, representing almost 10\% of the world's crude reserves and 13\% of reserves held by the Organization of Petroleum Exporting Countries 
 \cite{FT:2019}.
  The Petroleum industry in Iran accounted for 60\% of total government 
  revenues and 80 \% of the total annual value of both exports 
  and foreign currency earnings in 2009 \cite{Press.TV:2010}.
The oil price volatility should strongly affect the Tehran Stock Exchange index (TSE).

Gold is one of the basic assets ensuring stability of national currency. 
In an economic downturn, people tend more towards gold as a safe asset
while reducing their investments that   
 leads to deeper economic downturn
\cite{Nemat:2016}.
Iran's gold bar and coin sales tripled to 15.2 tons in the second quarter of 2018, the highest in four years as announced by the World Gold Council. 
According to Bloomberg, Iran's demand for gold bars and coins may remain strong for the rest of the year and even increase as the US reimposes sanctions
\cite{FT:2018}.
The gold price volatility should strongly affect the Tehran Stock Exchange index (TSE).
    
\begin{figure}[ht!]
	\centering
	\includegraphics[width=1\textwidth]{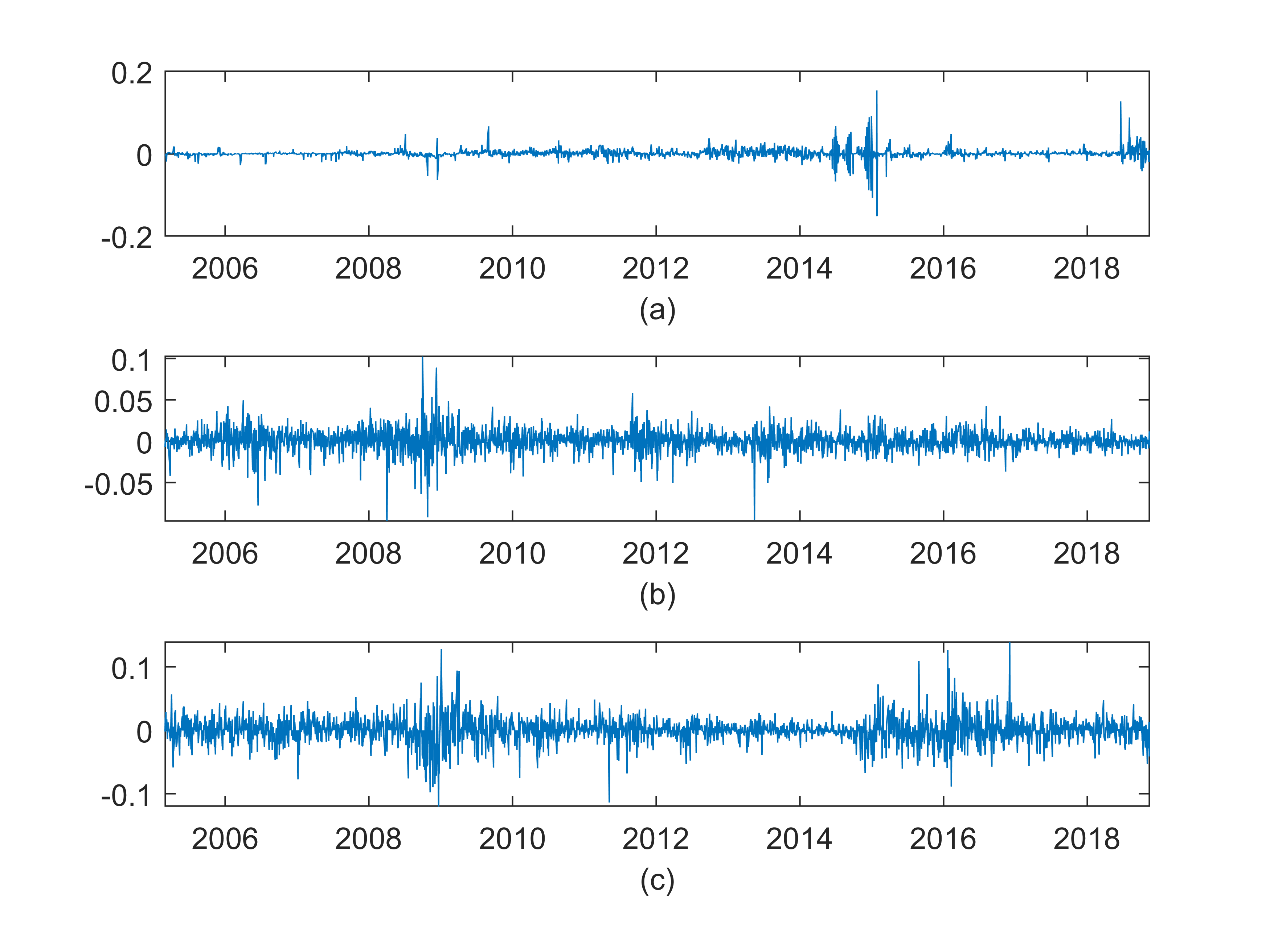}
	\caption{Daily log-returns for the period from 
		28/02/2005 to 14/11/2018  for the (a.) Tehran Stock Exchange; (b.) Gold; (c.) Crude Oil price.}
	\label{Log_return}
\end{figure}
 
The main objective of this paper is to estimate stochastic dependence 
between the daily log-returns for oil and gold and TSE 
by applying the copula method
for constructing non-Gaussian multivariate distributions and understanding the relationships among multivariate data.
We also assess the degree of dependence between extreme events
 on the national oil and gold markets and TSE.
   Although the linear correlation analysis shows 
  positive correlations between the quantities of interest, 
  the copula method indicates that the degree of dependence 
  between oil and TSE 
  is weak while there might be 
  a significant left tail dependence between TSE and gold
  that can be thought of as a proxy for systemic risk of default. 
  
  We use daily data for the closing Euro Brent Crude-oil price (in US dollars per share) and closing gold historical spot price (in US dollars per ounce) obtained from 
  the Historical London Fix Prices \cite{LFP}.   
The daily adjusted closing price for the TSE were taken from the official web site of the Tehran Stock Exchange \cite{TSE}. The time stamps of TSE prices were converted from the Persian to Western dates. 
 The daily log-returns of data time series (for gold, oil, and TSE) for the period from 
  28/02/2005 to 14/11/2018 are plotted in  Fig.~\ref{Log_return}.

  In Sec. 2, we provide a literature review 
   on the tail dependence between oil, gold, and stock market indices
   in Iran and worldwide.
  In Sec. 3, we give a brief introduction to the copula method.
  Our main contribution is explained in Sec. 4. 
  We apply the tail copula method to investigate 
  the dependence of joint extreme events for the gold and oil prices and TSE index.
  The standard ARMA-GARCH model is used to estimate the marginal distributions and to filter out the serial dependence and volatility clustering in the data. 
  Several copula models are implemented to the standardized residuals of each series.
  We estimate and model the tail dependence for the gold and oil prices and TSE index empirically and theoretically.   
    We then 
  conclude in the last section.   

\section{Literature review}

The impact of crude oil and gold prices on financial markets 
has been broadly discussed in the literature.  
 Hamilton (1983) had shown that the 
 oil price has an essential impact on the economy and its volatility 
 leads to 
 a stock market price change \cite{Hamilton:1983}.
  Kling (1985) applied the Vector Autoregressive (VAR) model to assess the impact of oil price movement on the S\&P 500 Index and
   five US industries \cite {Kling:1985}. 
   Huang et al. (1996) considered the relationship between log-return of oil and US stock market, by using VAR model \cite {Huang:1996} and demonstrated that the oil futures might affect  the stock returns of oil companies. Sadorsky (1996) used the unrestricted VAR model with generalized autoregressive conditional heteroskedasticity (GARCH) model to show the negative impact of increasing oil prices on the US stock market \cite {Sadorsky:1999}.

Cai et al. (2001) observed that the GDP per capita and the inflation rate have a substantial effect on the gold price volatility \cite {Cai:2011}. Baryshevsky (2004) observed  the high inverse correlation between real stock returns and the ten years average rate of gold \cite {Baryshevsky:2004}. Nandha and Faff (2008) reported 
the  strong correlation between oil prices and the stock market in oil exporting countries
\cite{Nandha:2008}.

Miller and Ratti (2009) found that after 1994 the long-term response of
market indices to the oil price shocks was negative \cite{Miller:2009}.
Baur and Mcdermott (2010)
studied the impact of gold price on the financial market during 1979-2009 and
 observed that gold acts as a hedge in the stock market of the US and most European countries \cite{Baur:2010}. The empirical findings of Batten et al. (2010) was that volatility of the gold return has a substantial impact on financial market 
 returns \cite{Batten:2010}.

Recently, several authors 
have applied time series models and the copula method to 
measure co-movements in the tails of the 
multivariate distributions.
 Bharn and Nikolovann (2010) considered the relationship between oil price and stock markets in Russia by applying the exponential GARCH model \cite {Bharn:2010}. 
 Fillis et al. (2010) studied the time-varying dependence between oil price and stock market in different countries during 1997-2009 by applying the DCC-MGARCH model \cite{Filis:2011}.
   Arouri et al. (2011) measured the degree of dependence between the oil price shocks and the Gulf Cooperation Council (GCC) stock market by the VAR-GARCH model \cite {Arouri:2011b}. J\"{a}schke et al. (2012) modeled large co-movements of the commodity returns by applying the copula method to extreme events on the energy market \cite{Jaschke:2012}.

Applying the ARJI-GARCH model, Chang (2012) observed that the tail dependence between crude oil spot and the future market is time-varying and asymmetric
\cite{Chang:2012}.  
With the use of  Archimedean copulas, Nguyen and Bhatti (2012)
 found no significant evidence
 of tail dependence between the stock market indices and oil price changes 
 in China and Vietnam \cite {Nguyen:2012}.
 Aloui et al. (2013) implemented three Archimedian copulas to measure
  the significant impact of crude oil price changes on the stock market indices of  Gulf Cooperation Council economies \cite {Aloui:2013}.

  Mensi et al. (2013) studied volatility transmission across the gold, oil, and equity markets and demonstrated the effect the gold and oil price variations have on the S\&P 500 index \cite {Mensi:2013}. 
 The copula method had been used  for exploring the tail dependence 
 between the financial and credit default swap markets by Silva et al. (2014) \cite{Silva:2014}. Arouri et al. (2014)  observed the significant impact of gold price volatility on the Chinese stock market in 2004 - 2011  by applying the VAR-GARCH model
\cite{Arouri:2014}.
Using the quantile regression method, Zhu et al. (2016) found 
the strong dependence at the upper and lower tails between the crude oil prices 
and the  Asia-pacific stock market index for 2000-2016 \cite {Zhu:2016b}.
   Siami-Namini and Hudson (2017) examined the volatility spillover from the returns 
   on crude oil to the commodities returns by using the Autoregressive (AR) model with the Exponential GARCH model \cite {sima:2017}, in the period from Jan 2006 to Nov 2015. 
      Trabelsi (2017)  studied the asymmetric tail dependence between international oil market and the Saudi Arabia sector indices in 2007-2016 \cite {Trabelsi:2016b}. 
         Hamma et al. (2018) applied the copula method and ARMA-GARCH-GDD model to analyze the dependence between  stock market indices
         of Tunisia and Egypt and the crude oil price in 1998-2013 \cite {Hamma:2018}.

Few scholars have investigated 
the dependence between the oil and gold prices and the stock market index in Iran. Foster and Kharazi (2008) found no significant correlation 
between the variations of oil price and TSE in 1997-2002 \cite {Foster:2008}. 
Applying the ARMA-copula method, 
Najafabadi et al. (2012) observed  
that the gold and oil price changes might weakly influence TSE
\cite{Najafabadi:2012}.  
 Shams and Zarshenas (2014) 
 agreed on that there is no significant evidence of dependence
  between the oil and gold price variations and the TSE index \cite{Shams:2014}.

  \section{Methods}

We investigate time series of 
the daily log-returns, 
\begin{equation}
\label{log_ret}
r_t= \log {\left( \frac{S_t}{S_{t-1}}\right) } ,
\end{equation}
for the closing prices of an asset $S_t.$ 

\subsection{Fitting the time series by the ARMA -- GARCH model}

The ARMA-GARCH model \cite{Fuller:1976, Hamilton:1994} is 
a standard tool for modeling the conditional mean and volatility of 
 time series. The GARCH model captures several important 
 characteristics of financial time series, including
   the heavy tail distribution of returns and volatility clustering. 
   The ARMA-GARCH model 
   filters  out the linear and nonlinear temporal dependence in bi-variate times series,
   \begin{equation}
\label{arma-garch1}
\begin{array}{lcl}
r_t &= &\mu+ \sum_{t=i}^{p} \varphi _i \, \left( r_{t-i}-\mu \right) +
\sum_{j=1}^{q} \theta _j \,  a_{t-j} +a_t,\\
a_t &=&\varepsilon_t \sigma_t, \,\,\,  \varepsilon_t \sim \mathrm{iid,} \\
\sigma_t^2 &=&  \gamma+ \sum_{m=1}^{k} \beta _m \,  \sigma^2_{t-m}+
\sum_{n=1}^{l} \alpha _n \,  a^2_{t-n}.
\end{array}
\end{equation}
where $r_t$ is the assets return, $\mu$ in constant term, $p$ and $q$ are the lag orders of ARMA model, $k$ and $l$ are the lag orders of GACH model, $\sigma_t=var\left( r_t \mid F_{t-1}\right)$ is the conditional variance during the period $t$,   $F_{t-1}$ denotes the information set consisting of all linear functions of the past returns available during the time period $t-1$,
$\varepsilon_t$ is the standardized residual 
during the time period $t$, which are {\it iid} with zero mean and unit variance;
$a_t$ is referred  to the shock of return during the period $t$; $\gamma\geq 0$ is a constant term;
$\theta_j \left\lbrace j=1,2,...,q\right\rbrace$ ,$\varphi_i \left\lbrace i=1,2,...,p\right\rbrace$, $\alpha_n \geq 0 \left\lbrace n=1,2,...,l\right\rbrace$ and $ \beta_m\geq 0 \left\lbrace m=1,2,...,k\right\rbrace$ are the parameters of the model estimated from the data.
 
\subsection{Tail copula method}

A copula is a multivariate probability distribution for which the marginal-probability distribution of each variable is uniform \cite {Sklar:1959}. 
Application of the copula method to description of the dependence between random variables in finance is relatively new.
 
In line with Sklar's Theorem \cite{Sklar:1959},  
every cumulative bivariate distribution $F$ with marginal 
distributions $F_1$ and $F_2$ can be written as
\begin{equation}
\label{copula_eq}
F\left(x_1,x_2\right)\,\,=\,\,C\left(F_1\left(x_1\right),F_2\left(x_2\right)\right),
\end{equation}
for some copula $C$, which is uniquely determined on 
the interval $[0,1]^2$. 

Conversely, any copula $C$ may be used to design a joint 
bivariate distribution $F$ from any pair of 
univariate distributions $F_1$ and $F_2$, 
viz., 

$$ C\left(u,v\right)\,\,=\,\,F\left(F_1^{-1}\left(u\right),F_2^{-1}\left(v\right)\right),$$

where $F_1^{-1}$ and $F_2^{-1}$ are the quantile functions of the respective marginal distributions.
If we have a random vector $X=(X_1,X_2)$, 
the copula for their joint distribution is
\begin{equation}
\label{copula_equation}
C\left(u,v\right)\,\,=\,\,P\left(U\leq u,V\leq v \right)=
F\left(F_1^{-1}\left(u\right),F_2^{-1}\left(v \right)\right),
\end{equation}
for all $u,v \in [0,1]$. 

We also use the survival copula 
$\bar{C}$
that links the joint survival function
$\bar {F}\left(x\right)=\,\,1-F\left(x\right)$
to the univariate marginal distribution, viz.,
\begin{equation}
\label{copula_equ}
\bar {C}\left(u,v\right)\,\,=\,\,\bar{C}\left(\bar {F_1}\left(u\right),\bar {F_2}\left(v\right)\right)=\,\,\Pr\left(X_1\geq u,X_2\geq v \right).
\end{equation}
Following \cite{Nelsen:1999}, we can write survival copula function as  
\begin{equation}
\label{copula_Su}
\bar {C}\left(u,v\right)\,\,=\,\,\Pr\left(U\geq {1-u},V\geq {1-v} \right)=
\bar{F}\left(F_1^{-1}\left(1-u\right),F_2^{-1}\left(1-v\right)\right).
\end{equation}

\subsection{Rank correlation and tail dependence coefficients}

The Kendall and Spearman rank correlation coefficients 
of two variables,  $X_1$ and $X_2,$
with 
the copula 
$C\left(u,v\right)$ 
 are given by 
\begin{equation}
\label{Kendall}
\tau \left(X_1,X_2)\right.= 4 \int_{0}^{1} \int_{0}^{1} C \left(u,v)\right. dC \left(u,v)\right. -1, 
\end{equation}
and 
\begin{equation}
\label{Spearman}
\rho_S \left(X_1,X_2)\right.= 12 \int_{0}^{1} \int_{0}^{1} C \left(u,v)\right. du\,\,dv \,-3, 
\end{equation}
respectively. 

Let the distribution of $X_1$ and $X_2$ denote by $F_1$ and $F_2$. 
The following relation exists between Spearman's rank and linear correlation coefficients: 
\begin{equation}
\label{Linear_corr}
\rho_{s} \left(X_1,X_2)\right.=\,\,C\left(F_1\left(x_1\right),F_2\left(x_2\right)\right), 
\end{equation}
where $\left(u,v)\right.=\,\,\left(F_1(x),F_2(x)\right)$.\\
The relations between Kendall's Tau and Spearman's rank correlation
coefficients and the coefficient of linear correlations
in the Gaussian and Student's t-copulas are 
\begin{equation}
\label{corr-sper}
Corr(X_1,X_2)\,\,=\,\,\sin\left(\frac{\pi}{2} {\tau}\right),
\end{equation}
\begin{equation}
\label{corr-Kend}
{Corr}(X_1,X_2)\,\,=\,\,\sin\left(\frac{\pi}{6} {\rho_{s}}\right).
\end{equation}
Both $\rho_{s}$ and $\tau$ may be considered as a measure of the degree of monotone dependence between random variables, whereas the linear correlation coefficient measures the degree of linear dependence only. Since $\tau$ and $\rho_{s}$   measure the dependence in centered data, 
they are often insufficient to estimate and describe the dependence structure of extreme events. 
Hence, according to Embrechts et al. (1999) \cite{Embrechts:1999}, it is significantly better to use 
the tail copula method than the linear correlation coefficient to characterize the dependence of extreme events. In their opinion, one should choose a model for the dependence structure that reflects more detailed knowledge of the value at risk, and 
the tail copula method 
is an excellent tool for managing risks against concurrent events.

The standard way to assess tail dependence is to look at the lower and upper tail coefficients 
denoted by $\lambda_l $ and $\lambda_u$, respectively, where  
$\lambda_u$  quantifies the probability to observe a large $X_1$ value given the large value of $X_2$. Similarly, $\lambda_l$  is a measure that quantifies the probability to observe a small  $X_1$ value, assuming that the value of $X_2$ is small. 
Let $\,X_i\sim {F_{X_i}}$ and 
the probability  
$\alpha \in (0,1)$ 
 then the upper tail coefficient is:
\begin{equation}
\label{upper-tail}
\lambda_{u}\left(X_1,X_2)\right.=\lim_{\alpha \to 1} \,\,\Pr\left(X_{1}\geq F_1^{-1} \left(\alpha \right) \mid \, X_2 \geq F_2^{-1}\left(\alpha  \right)\right),
\end{equation}
and similarly  
\begin{equation}
\label{Lower-tail}
\lambda_{l}\left(X_1,X_2)\right.=\lim_{\alpha \to 0} \,\,\Pr\left(X_{1}\leq F_1^{-1} \left(\alpha \right) \mid \, X_2 \leq F_2^{-1}\left(\alpha  \right)\right).
\end{equation}
On the one hand should $\lambda_{l}\left(X_1,X_2)\right.=0$ $(\lambda_{u}\left(X_1,X_2)\right.=0)$ then  $X_1$ and $X_2$ are said to be lower (upper) asymptotically independent. On the other hand should  $\lambda_{l}\left(X_1,X_2)\right.>0$ $(\lambda_{u}\left(X_1,X_2)\right.>0)$, then the small (large) events tend to happen coherently, and  $X_1$ and $X_2$ are lower (upper) tail dependent. 

According to  J\"{a}schke (2012) \cite {Jaschke:2012}, the Value-at-Risk (VaR) is closely related to the concept of tail dependence. Tail dependencies can be considered as the degree of likelihood of an asset return falling below its VaR at the certain level $\alpha$ when the other asset returns have fallen below its VaR at the same level.
In general, $\lambda_{l}$ and $ \lambda_{u}$ (like the other scalar quantities) describe 
a certain level of dependence in tails. However, in our analysis, we need to describe the general framework of tail dependence for bi-variate distributions. 

In the general framework of tail copulas, the dependence structure of extreme events in bi-variate distributions, independently of their marginals, is represented by tail dependence (see Schimtz and Stadtm\"{u}ller (2006) \cite { Schmidt:2006} for more details). 

The lower and upper tail dependence
 associated with $X_1$ and $X_2$ are
\begin{equation}
\label{lower-copula}
\Lambda_{L}\left(X_1,X_2)\right.=\lim_{t \to \infty} \,t\, C\left(\frac{x_{1}}{t} ,\frac {x_2}{t} \right),
\end{equation}
\begin{equation}
\label{upper-copula}
\Lambda_{U}\left(X_1,X_2)\right.=\lim_{t \to \infty} \,t\, \bar {C}\left(\frac{x_{1}}{t} ,\frac {x_2}{t} \right)
\end{equation}
provided the above limits exist everywhere on $ {\rm I\!R}_{+}^2 := [0,\infty)^2.$ 
According to Schimtz and Stadtm\"{u}ller (2006) \cite {Schmidt:2006}, the estimation of tail dependence is a nontrivial task, particularly for a non-standard distribution
that is why we consider the tail coefficient as a measure of the tail dependence. 
The tail coefficient is a specific case of tail dependence, and we have  $\lambda_l = \Lambda_{L}\left(1,1)\right.$ and $\lambda_u = \Lambda_{U}\left(1,1)\right.$.
 
\section{Results}

\subsection{Model selection}

First, we perform the Ljung-Box $Q$-test \cite{Ljung:1978} to examine the presence of autocorrelation in log-returns and the presence of heteroskedasticity in squared log-return of 
each data set for lags 5 and 10.
 The $p$-values $\left( p<0.01\right)$ indicate that  
 the log-return of data
 (on gold, oil, and TSE) 
  are  autocorrelated in each times series.
 There are significant heteroskedasticity effects,
since all $p$-values are less than 0.01.
 The hypothesis of independent and identically distributed data is therefore rejected.
 
 Second, to investigate the multiple co-integration relationships among the gold, crude oil, and TSE
 time series,
 we apply the Engle-Granger co-integration test \cite{Fuller:1976, Hamilton:1994}.
 The obtained $p$-values $( p_1=0.0102,\,\,p_2=0.0000)$ reject the null hypothesis 
 of no co-integration among the time series. 
 The test result indicates that there is a long-run relationship 
 among the variables, and they share a common stochastic drift. 
 
 Third, the linear and nonlinear temporal dependencies in bi-variate times series should be filtered
  out by applying the ARMA-GARCH  model \cite{Fuller:1976, Hamilton:1994}.
  We tried the ARMA-GRACH model with various lags and different distributions 
   to select the optimal model for each times series. Namely, we have tested the normal, Student's t, generalized error, skewed Student's t, and generalized hyperbolic distributions. 
  
\begin{table}[b!]
	\centering
	\caption{Long memory test in TSE, Oil, and Gold log-return}
	\begin{tabular}{@{}lcc@{}}
		\toprule
		\multicolumn{1}{p{7.215em}}{Data} & \multicolumn{2}{c}{FARIMA(1,$d$,1)-GARCH(1,1)} \\ \midrule
		& $d$                     & $p$-value              \\
		TSE                       & 0.142              & 0.000                    \\
		Oil                      & 0.042              & 0.059             \\
		Gold                      & 0.050              & 0.102            \\ \bottomrule
	\end{tabular}
\label{table:Fractional}
\end{table}

To examine the presence of long memory dependence in each times series,
 we perform the fractional ARMA(1,d,1)-GARCH(1,1) test.
 The results of all tests  are given in 
  Tab.~\ref{table:Fractional}.
  
  The test $p$-values for gold and oil log-returns $\left(p\geq0.05\right)$) demonstrate {\it no long memory dependence} in oil and gold times series. However,
  the very small $p$-value for TSE log-return $ \left( p= 0.000\right)$
  shows that there is a long memory dependence in the TSE time series. 
  The fractional exponent 
  $\left(d\simeq0.142 \right)$ of the shift operator $(1-B)$ 
   confirms the presence of long memory dependence in the TSE log-returns.

Fourth, we have evaluated all possible ARMA-GARCH models ($p \leq 2, q\leq 2,k\leq 2,l\leq 2$)
with the use of the R-Package \textit {"rugarch"} \cite{Ghalanos:2018} by examining 
i.) Akaike Information Criterion (AIC); ii.) Bayesian Information Criterion (BIC) \cite{Fuller:1976, Hamilton:1994}; iii.) the statistical
significance test of model parameters at the $5\%$ -level; 
iv.)  lack of autocorrelation; v.) lack of heteroskedasticity 
in standardized residuals of each time series for lags 5 and 10.

Fifth, to evaluate the density of standardized residuals (innovations) in each 
time series, we apply probability integral transform method
following \cite{Diebold:1998}. 
This method is based on the relation between the sequence of 
densities of the standardized residuals, $p_t\left( z_t\right),$ 
and its integral probability transform,  
\begin{equation}
\label{probability_integral}
y_t=\int_{-\infty}^{z_t}  p_t \left(u)\right. du. 
\end{equation}
We evaluate the densities of the standardized residuals 
by assessing whether the probability integral transform series,
$\left\lbrace y_t\right\rbrace _{t=1}^{m},$ are $iid\,\, U(0,1)$. 
The non-parametric Kolmogorov-Smirnov test  is the easiest way 
to check the uniformity of $y_t$'s \cite{Kolmogorov:1933}.

By comparing all criteria and evaluating
the densities of the standardized residuals, 
the obtained time series models 
are 
the FARIMA $\left( 1, 0.14 , 0\right)$ - GARCH $\left( 1, 1\right)$ model with 
the Standardized Generalized Hyperbolic Distribution (SGHYD),
 the ARMA $\left(1,1\right)-$ GARCH $\left(1,1\right)$ model with Student's t-distribution with $v=3.363$ degree of freedom, 
 and the GARCH$\left(1,1\right)$  model with SGHYD, 
 for TSE, gold, and oil time series, respectively. 
  \begin{table}[ht!]
	\centering
	\caption{Maxmium likelihood estimation, standard errors, and estimations of model parameters in FARMA$(1,0)$-GARCH$(1,1)$  with SGHYD for TSE, the ARMA$(1,1)$-GARCH$(1,1)$ model with Student's t-distribution for oil, and the GARCH$(1,1)$  model with SGHYD for gold.}
	\begin{tabular}[b!]{lrcccccccc}
		\toprule
		\textbf{Parameter} &       & \multicolumn{2}{c}{TSE log-return} &       & \multicolumn{2}{c}{Oil log-return} &       & \multicolumn{2}{c}{Gold log-return} \\
		\cmidrule{3-4}\cmidrule{6-7}\cmidrule{9-10}          &       & Esitmation & Std. error &       & Esitmation & Std. error &       & Esitmation & Std. error \\
		\midrule
		\textbf{ARMA-GARCH model} &       &       &       &       &       &       &       &       &  \\
		$\varphi_1$   &       & 0.1477 & 0.0273 &       & 0.8705 & 0.2250 &       &  ---     & --- \\
		$\theta_1$ &       & ---   & ---   &       & -0.7850 & 0.2239 &       &   ---    & --- \\
		$\alpha_1$ &       & 0.3616 & 0.0277 &       & 0.0348 & 0.0037 &       & 0.0285 & 0.0022 \\
		$\beta_1$ &       & 0.6374 & 0.0441 &       & 0.9642 & 0.0021 &       & 0.9655 & 0.0019 \\
		$d$-(arfima) &       & 0.1331 & 0.0203 &       & ---   & ---   &       &       &  \\
		\midrule
		\textbf{Distribution} &       &       &       &       &       &       &       &       &  \\
		$\nu\,$(DF)    &       & 0.2500 & 0.0158 &       & 3.3629 & 0.2743 &       & 0.2500 & 0.0253 \\
		$\zeta\,$  &       & -0.0327 & 0.0613 &       & ---   & ---   &       & 0.0035 & 0.0187 \\
		$\eta\,$  &       & -1.1811 & 0.1183 &       & ---   & ---   &       & 0.3665 & 0.0657 \\
		\bottomrule
	\end{tabular}%
	\label{tab:model estimation}%
\end{table}%
The estimated parameters of each model 
obtained by using the maximum likelihood methods
are given in 
Tab.~\ref{tab:model estimation}.  

\begin{figure}[ht!]
	\centering
	\includegraphics[width=1\textwidth]{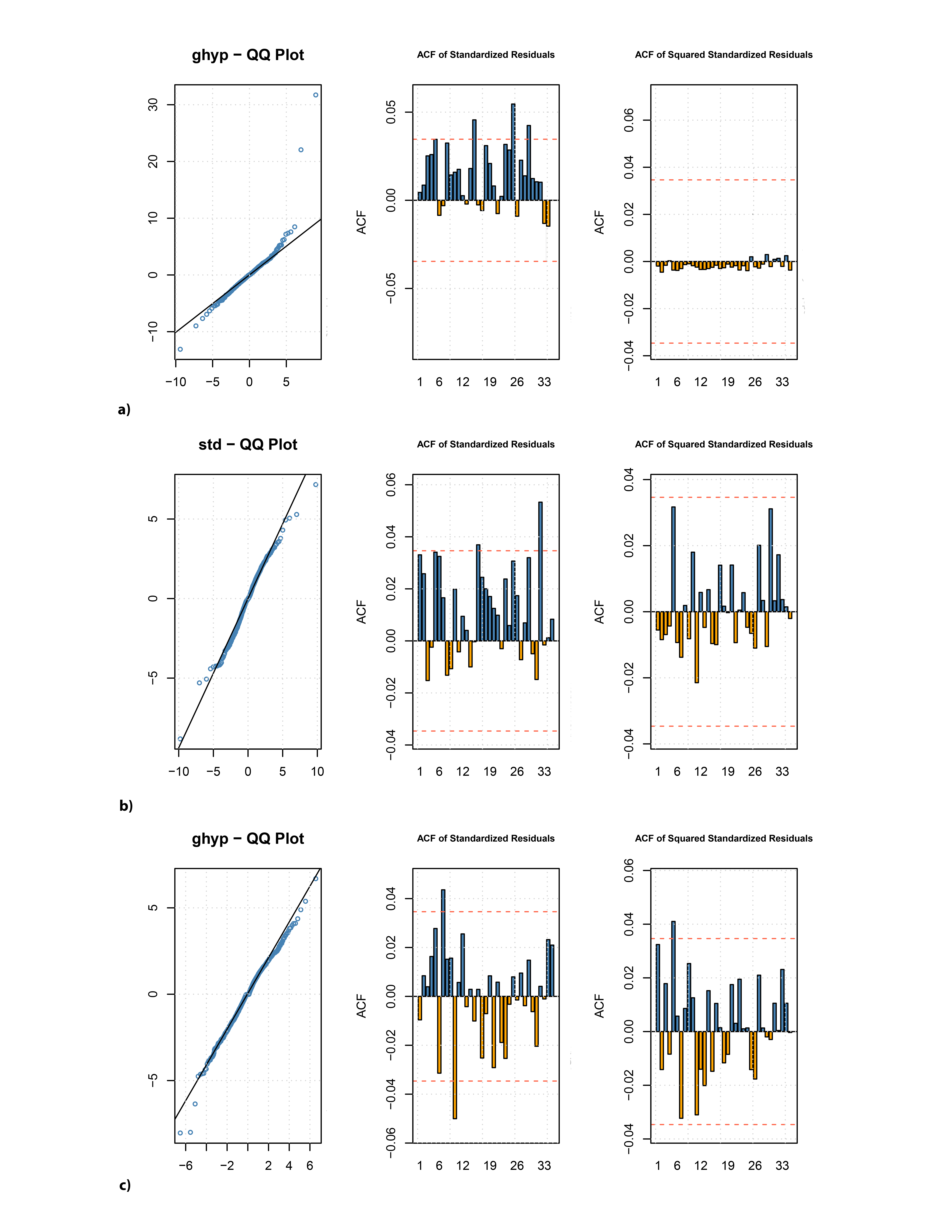}
	\caption{QQ plot and correlogram of  ACF of standardized residuals and ACF of squared  standardized residuals for (a). TSE log-return, (b). Oil log-return, (C) Gold log-return}
	\label{oil_plot}
\end{figure}

 To study autocorrelation and conditional heteroskedasticity in each residual time series, 
 we have shown the QQ plots, and correlograms for $z_t$ and $z_t^2$, for each standardized residual series
 in Fig.~\ref{oil_plot}.
The apparent linearity of the QQ-plots 
shows that the corresponding distributions are well-fitted. 

The parameters for each model were estimated by the maximum-likelihood method
and all parameters are found significant at the 5\% -level.
  Since the $p$-values in the Ljung-Box Q-tests are less than $0.01$,
   there is no significant autocorrelation and heteroskedasticity of the standardized residuals
  at the $5\% $-level for each time series.
    
For evaluating the forecast densities of each model, 
first we calculate the $\left\lbrace y_t\right\rbrace _{t=1}^{m}$ for each standardized residual sets associated with gold, oil and TSE. 
Then, we test that $y_t$'s are $iid\,\, U(0,1)$ by the
 Kolmogorov-Smirnov and Sarkadi-Kosik tests \cite{Kosik:1985} with the use of 
 the R-package \textit {"uniftest"} \cite{uniftest:2015}.
  The $p$-values ($p\geq 0.05$) indicate that  
  $y_t$'s are $iid\,\, U(0,1)$ at the 5\%-level.

 We conclude that the obtained distributions are well fitted.

Therefore, the optimal ARMA-GARCH model for the oil log-returns is 
\begin{equation}
\label{arma_oil}
\begin{array}{lcl}
O_t=\,0.000343+0.870548\,O_{t-1}\,-\,0.784959\,a_{t-1}\,+\,a_t, \\
a_t=o_t\,\sigma_t\,,\,\,\,\, o_t\,\, iid \,\, \sim \,\,t_{\left( \nu=3.363\right)} \,,\\
\sigma_t^2\,=\,0.034843\,o_{t-1}^2\,+\,0.954157\,\sigma_{t-1}^2 \,.    
\end{array}
\end{equation}
The best ARMA-GARCH model for the gold log-returns is
\begin{equation}
\label{arma_gold}
\begin{array}{lcl}
G_t=0.000175\,+\,a_t, \\
a_t=g_t\,\sigma_t\,,\,\,\,\, g_t\,\, iid \,\, \sim\,\,SGHYD\left( 0.00348,0.2500 \right),\\
\sigma_t^2\,=\,0.028526\,g_{t-1}^2\,+\,0.945546\,\sigma_{t-1}^2  \,.   
\end{array}
\end{equation}
Finally, 
the optimal model for the log-returns for TSE is 
\begin{equation}
\label{arma_TSE}
\begin{array}{lcl}
(1-B)^{0.14}S_t=\,0.000251+0.147724\,S_{t-1}\,+\,a_t, \\
a_t=s_t\,\sigma_t\,,\,\,\,\, s_t\,\, iid \,\, \sim\,\,SGHYD\left(-0.032714,0.2501 \right),\\
\sigma_t^2\,=\,0.361576\,s_{t-1}^2\,+\,0.627424\,\sigma_{t-1}^2 \,.    
\end{array}
\end{equation}
where $B$ is back-shift operator. The \textit{rugarch} \cite{Ghalanos:2018} package performs GHYD parameter estimations using the $(\eta,\zeta)$ parametrization (SGHYD), after which a series of steps transform those parameters into the $(\mu,\alpha, \beta, \delta )$ while at the same time including the necessary recursive substitution of parameters in order to standardize the resulting distribution.

 \subsection{Tail dependence measure }

\subsubsection{Selection of the copula model}

In this section, we model the dependence structure 
between  the log-returns of oil, gold, and TSE denoted by $O_t$, $G_t $, and $S_t$, respectively, by the copula method.  
We have already shown that the  $O_t$, $G_t$ and $S_t$ 
are time-varying dependence series.
We applied the time series model to filter out time-varying dependence and obtain standardized residuals for gold, oil,  and TSE, denoted by $g_t$, $o_t$, and $s_t$, respectively. 
 We work with the iid standardized residuals for each time series.

Deheuvels et al. (1979) \cite {Deheuvels:1979} introduced the empirical copula based on the rank correlation. An initial approach to calculate the empirical copula function is to estimate the distribution function of residuals based on 
the empirical distribution function $F_X$, 
\begin{equation}
\label{pseudo}
F_X \left(x \right) = \frac{1}{T+1}  \sum_{i=1}^{T} \mathds{1}  \left( X_i \leq x \right),
\end{equation}
where $\mathds{1}(A)$ denotes the indicator function of the set $A$.
The quantities $F_X \left( x_i \right) $ and $F_Y\left( y_i\right)$ 
as given by (\ref{pseudo}) are the ranks of 
$X_i$ and $Y_i$ normalized  by $T+1$. 

Residual sets $o_t$, $g_t$, and $s_t$ are transformed to the rank-based variables
by  
\begin{equation}
\label{rank_var}
u_t= \frac{rank \left( o_t )\right.}{T+1}, \,\,\,\,\, v_t= \frac{rank \left( g_t)\right. }{T+1}, \,\,\,\,\, w_t= \frac{rank \left( s_t )\right. }{T+1}. \,\,\,\,\, 
\end{equation}
The domain of empirical copula, $\left( u,v\right)\in \left[ 0,1\right] ^2$ \cite{Genest:2007},
is formed by all  normalized ranks, $\left\lbrace \frac{1}{T+1},\frac{2}{T+1},...,\frac{T}{T+1}\right\rbrace,$ for each residual set.
The best sample-based empirical copula is then defined by 
\begin{equation}
\label{em_copula}
C_u \left(u,v \right) = \frac{1}{T}  \sum_{t=1}^{T} \mathds{1}  \left( U_t < u, V_t < v \right),
\end{equation}
and the empirical survival copula is 
\begin{equation}
\label{sur_copula}
\bar {C}_u \left(u,v \right) = \frac{1}{T}  \sum_{t=1}^{T} \mathds{1}  \left( U_t \geq u, V_t \geq v \right).
\end{equation}
Before using the obtained copulas, 
we have estimated the Spearman  $\rho$ and Kendall $\tau$ rank correlation coefficients 
from the normalized ranks (see Tab.~\ref{tab:correlation}). The 
obtained coefficients show that 
the dependence between oil and TSE is very weak, but 
the correlation between gold and TSE is significant.
  \begin{table}[ht!]
  	\centering
  	\caption{Correlation between ranks of standardized residual of gold, oil and TSE}
  	\begin{tabular}{rcccrc}
  		\cmidrule{1-5}\cmidrule{5-6}    \multicolumn{1}{l}{Method} & \multicolumn{2}{c}{Kendall} &       & \multicolumn{2}{c}{Spearman} \\
  		\cmidrule{1-3}\cmidrule{5-6}          & \multicolumn{1}{c}{Gold                       } & \multicolumn{1}{c}{Oil} &       & \multicolumn{1}{c}{Gold                       } & \multicolumn{1}{c}{Oil} \\
  		\multicolumn{1}{l}{TSE                             } & \multicolumn{1}{r}{0.111} & \multicolumn{1}{r}{0.017} &       & 0.162 & \multicolumn{1}{r}{0.026} \\
  		&       &       &       &       &  \\
  		\cmidrule{1-5}\cmidrule{5-6}    \end{tabular}%
  	\label{tab:correlation}%
  \end{table}%
The dependence between gold, oil, and TSE was also assessed by applying the multivariate independent test \cite{Genest:2011}. The obtained $p$-value $\left( p=0.021\right)$ at the $5\%$ level rejects the hypothesis of independence of these variables. 
We conclude that the Spearman and Kendall's correlations
can not  measure the entire dependence 
between the quantities of interest.

Now we determine the optimal 
copula model to describe the dependence structure of the joint distributions. 
We consider both theoretical (parametric) $C$ and empirical (non-parametric) $C_u$ methods.
We fit several copula families [Independent, Gaussian, Student's t, Clayton, Gumbel, Frank and Joe] on standardized residuals. 
 The unknown copula parameter $  \theta  $ is 
obtained  by the {\it inverse $\tau$ method}, i.e., by
solving the equation (\ref{corr-Kend})
for the value $\hat \tau$ 
 estimated from the data with the use of the R-package 'VineCopula' \cite{vinecopula:2011}. 
 The parameter
   $\theta$ controls the strength of dependence in each copula family;
   its values are collected in Tab.~\ref{tab:copula}.
   The significance of each fitted copula is examined by the goodness of fit test with the use of Cramer-Von-Mises statistics \cite {Genest:2009}:
   \begin{equation}
\label{good_test}
 S_t = \sum_{t=1}^{T} \left( C_t\left( u_t,v_t\right) - C_{\hat \theta_t} \left( u_t,v_t\right)  \right)^2.
\end{equation}
The corresponding $p$-value is obtained by the bootstrapping method (see \cite {Genest:2008} for details). 
 We have evaluated each copula family with respect to 
 the following three criteria:
 i.) the high $p$-value of the goodness of fit test; 
 ii.) the AIC; and 
 ii.) the BIC values. 
  The obtained values are given in Tab.~\ref{tab:copula}.
 \begin{table}[htbp]
 	\centering
 	\caption{The copula parameters and goodness-of-test results for the different copula families.}
 	\begin{tabular}{rrlrrrllrrr}
 		\toprule
 		&       & \multicolumn{4}{c}{Oil and TSE} &       & \multicolumn{4}{c}{Gold and TSE} \\
 		\cmidrule{1-1}\cmidrule{2-8}\cmidrule{8-11}    \multicolumn{1}{l}{Copula } &       & \multicolumn{1}{c}{$\hat{\theta}$} & \multicolumn{1}{c}{AIC} & \multicolumn{1}{c}{BIC} & \multicolumn{1}{c}{$p$-value} &       & \multicolumn{1}{c}{$\hat{\theta}$} & \multicolumn{1}{c}{AIC} & \multicolumn{1}{c}{BIC} & \multicolumn{1}{c}{$p$-value} \\
 		\cmidrule{1-1}\cmidrule{3-6}\cmidrule{8-11}    \multicolumn{1}{l}{Independent} &       &       &       &       & 0.79  &          &       &       &       & 0.00 \\
 		\multicolumn{1}{l}{Gaussian} &       & \multicolumn{1}{r}{0.037} & -2.249 & 3.822 & 0.83  &  \,\,\,\,\,        & \multicolumn{1}{r}{0.174} & -82.875 & -76.803 & 0.02 \\
 		\multicolumn{1}{l}{t} &       & \multicolumn{1}{r}{0.034} & 6.544 & 18.688 & 0.04  &      \,\,\,\,\,   & \multicolumn{1}{r}{0.173} & -66.765 & -54.622 & 0.00 \\
 		\multicolumn{1}{l}{Clayton} &       & \multicolumn{1}{r}{0.028} & -0.451 & 5.621 & 0.63  &  \,\,\,\,\,        & \multicolumn{1}{r}{0.161} & -44.561 & -38.489 & 0.40 \\
 		\multicolumn{1}{l}{Gumbel} &       & \multicolumn{1}{r}{1.016} & -1.285 & 4.787 & 0.75  &   \,\,\,\,\,       & \multicolumn{1}{r}{1.080} & -50.448 & -44.377 & 0.07 \\
 		\multicolumn{1}{l}{Frank} &       & \multicolumn{1}{r}{0.151} & -0.053 & 6.019 & 0.34  &    \,\,\,\,\,      & \multicolumn{1}{r}{1.003} & -80.859 & -74.788 & 0.00 \\
 		\multicolumn{1}{l}{Joe} &       & \multicolumn{1}{r}{1.019} & -0.666 & 5.405 & 0.76  &    \,\,\,\,\,      & \multicolumn{1}{r}{1.080} & -19.940 & -13.868 & 0.63 \\
 		\cmidrule{1-1}\cmidrule{2-8}\cmidrule{8-11}          &       & a.)   &       &       &       &          & b.)   &       &       &  \\
 	\end{tabular}%
 	\label{tab:copula}%
 \end{table}%
  
We conclude that the Student's t copula is not appropriate for modeling the dependence between oil and TSE (as $p$-value$\leq 0.05$). The small goodness-of-fit $p$-values for gold and TSE 
also reject the Independent, Frank, and t copula families at the $5\%$ level. For any other candidates, the $p$-value can not reject the null hypothesis.

The Gaussian copula is a candidate for the optimal model describing the  
the dependence between oil and TSE, with the smallest AIC and BIC values, and the highest 
$p$-value among all other candidates. 

Due to zero $p$-value, the Frank copula is not a good fit for the description of dependence between gold and TSE although the values of  AIC $\left( -80.859\right) $  and BIC $\left(-74.788 \right) $ are the smallest ones.
A copula family with the non-trivial tail coefficients and higher $p$-value goes over
other families in explaining the entire dependence between gold and TSE. 
 Therefore, we have selected the
Joe copula as a proper candidate because of its higher $p$-value $\left(0.63 \right) $ 
 and the smaller AIC and BIC values over the Clayton copula  
 even though the lower tail dependence coefficient in Joe copula is zero. 
  
 \subsubsection{Lower and upper tail coefficients for oil, gold, and TSE}

J\"{a}schke et al. have shown that the goodness-of-fit test alone does not necessarily provide an appropriate model for tail dependence, because it is based on minimizing the distance between the observed ranks model 
and parametric model 
 over the whole support of the distribution \cite{Jaschke:2012}.
  They suggested applying the tail copulas concept for capturing dependence in the tail of distribution to improve the effectiveness of fitted copula. The non-parametric estimators
  are defined up to a scaling factor $k= 1,2,...,T $ chosen by a statistician \cite{Schmidt:2006},
   for the lower tail copula,  
   \begin{equation}
 \label{lower_copula}
\hat \Lambda_{L} \left( x,y\right) =\, \frac{T}{k} C_u \left(\frac{kx}{T},\frac{ky}{T} \right) = \frac{1}{k}  \sum_{t=1}^{T} \mathds{1}  \left( u_t \leq \frac{kx}{T+1} , v_t \leq \frac{ky}{T+1}\right),
\end{equation}
and for the upper tail copula,
\begin{equation}
\label{upper_copula}
\hat \Lambda_{U} \left( x,y\right) =\, \frac{T}{k} \bar {C}_u \left(\frac{kx}{T},\frac{ky}{T} \right) = \frac{1}{k}  \sum_{t=1}^{T} \mathds{1}  \left( u_t > \frac{T-kx}{T+1} , v_t > \frac{T-ky}{T+1}\right),
\end{equation}
respectively.
 Based on the above estimators, they found that 
\begin{equation}
\label{estimators}
\hat \lambda_l\,\, = \,\,\hat \Lambda_{L,T} \left( 1,1\right),
\quad \hat \lambda _ u\,\, =\,\, \hat \Lambda_{U,T} \left( 1,1\right)
\end{equation}
   are the appropriate non-parametric estimators for the upper and lower tail dependence coefficients.
 
 We estimate the lower and upper tail coefficients by using the equations
 (\ref{estimators}. 
  The values of coefficient estimators for tail dependence between oil and TSE, 
 gold and TSE
 are given in  
  Tab.~\ref{tab:tailcopula}). 
  \begin{table}[ht!]
  	\centering
  	\caption{Tail dependence coefficient estimators for tail dependence between oil \& TSE, 
  		gold \& TSE for the different copula families.}
  	\begin{tabular}{rrlrllr}
  		\toprule
  		&       & \multicolumn{2}{c}{Oil and TSE} &       & \multicolumn{2}{c}{Gold and TSE} \\
  		\cmidrule{1-1}\cmidrule{2-6}\cmidrule{6-7}    \multicolumn{1}{l}{Copula } &   \,\,\,\,\,\,\,\,\,\,\,\,    & \multicolumn{1}{c}{lower} & \multicolumn{1}{c}{upper} &     \,\,\,\,\,\,\,\,\,\,\,\,  & \multicolumn{1}{c}{lower} & \multicolumn{1}{c}{upper} \\
  		\cmidrule{1-1}\cmidrule{3-4}\cmidrule{6-7}    \multicolumn{1}{l}{Independent} &       & \multicolumn{1}{r}{0.0000} & 0.0000 &          & \multicolumn{1}{r}{0.0000} & 0.0000 \\
  		\multicolumn{1}{l}{Gaussian} &       & \multicolumn{1}{r}{0.0000} & 0.0000 &          & \multicolumn{1}{r}{0.0000} & 0.0000 \\
  		\multicolumn{1}{l}{Clayton} &       & \multicolumn{1}{r}{0.0000} & 0.0000 &       & \multicolumn{1}{r}{0.0434} & 0.0000 \\
  		\multicolumn{1}{l}{Gumbel} &       & \multicolumn{1}{r}{0.0000} & 0.0216 &          & \multicolumn{1}{r}{0.0000} & 0.1000 \\
  		\multicolumn{1}{l}{Frank} &       & \multicolumn{1}{r}{0.0000} & 0.0000 &          & \multicolumn{1}{r}{0.0000} & 0.0000 \\
  		\multicolumn{1}{l}{Joe} &       & \multicolumn{1}{r}{0.0000} & 0.0250 &          & \multicolumn{1}{r}{0.0000} & 0.1005 \\
  		\multicolumn{1}{l}{Empirical} &       & \multicolumn{1}{r}{0.0000} & 0.0000 &          & \multicolumn{1}{r}{0.0680} & 0.0000 \\
  		\cmidrule{1-1}\cmidrule{2-6}\cmidrule{6-7}          &       &    &       &          &    &  \\
  	\end{tabular}%
  	\label{tab:tailcopula}%
  \end{table}%
In the preceding section, our results based on the highest $p$-value, the smallest AIC and BIC values have suggested that the Gaussian copula is the optimal one for describing the dependence between oil and TSE. 
However, the zero values of estimators given in  
Tab.~\ref{tab:tailcopula} indicate that there is no significant 
tail dependence between oil and TSE, in line with the small values of the linear 
correlation coefficients given in Tab.~\ref{tab:correlation}. 
Therefore, we conclude that the independent copula
is the optimal model for describing the dependence between oil and TSE on the Iranian market.

Concerning the  tail dependence between gold and TSE, 
the results of the previous subsection show that the Joe copula 
is optimal for 
describing the dependence structure between gold and TSE. 
The non-parametric estimators for the 
upper and lower tail dependence between gold and TSE 
given in Tab.~\ref{tab:tailcopula}.
By comparing the empirical coefficients 
with the parametric tail coefficient of Joe copula,
we conclude that the Joe copula, which has the nontrivial 
upper tail and zero lower tail dependence, 
can not well enough explain the risk of extreme events in the tails.

From Tab.~\ref{tab:tailcopula}, we see that the
Gumbel copula goes over other copula families in explaining 
the dependence structure between gold and TSE
due to the small AIC and BIC values.
However, comparing the empirical coefficients 
  with the parametric tail coefficients of 
Gumbel copula, we see that the former is inverse with respect to the 
 Gumbel ones. Therefore, 
  we should reject the  Gumbel copula as well. 

We conclude that 
 the Clayton copula is the optimal model for describing the dependence structure 
 between gold and TSE on the Iranian market, because the lower tail coefficient 
 is close to empirical while the upper tail is zero. 
 The Clayton copula has the second highest $p$-value among the other candidates, 
 and its AIC and BIC values are small comparing to those for others copulas
 (see Tab.~\ref{tab:tailcopula}).

\section{Discussion and Conclusion}

{\bf Tails matter!} In our work, we convincingly demonstrate 
that the standard goodness measures for fitting copulas to data,
 such as the $p$-value, AIC/ BIC values,
 are not the reliable indicators of goodness-of-fit.
The tail dependence should always be taken into account 
as a proxy for systemic risk by risk managers.

Although the small AIC/ BIC values and the goodness-of-fit test
assume that the Gaussian copula is the good one 
for representing the dependence structure between oil and TSE 
on the Iranian market, the absence of tail dependence and low correlation between 
 these assets  disclose that the 
 independent copula is the optimal one for fitting the data. 
 Again, the small AIC/ BIC values, the goodness-of-fit test,
and the high $p$-value suggest that the Joe copula would be 
a good data model describing the relation between gold and TSE 
on the Iranian market.
However, our finding reveals that 
the Joe copula, which has only an upper tail and no lower tail dependence,
fails to describe the tail dependence of gold and TSE properly.
The empirical tail coefficients suggest that the Clayton copula
might be a suitable model fitting the data on gold and TSE well.

{\bf Watch the gold price!} Our finding has the important implications for  risk managers and investors,  because of
 it can help them to adjust the structure of their investment 
 portfolios once the gold price changes.
 Risk managers should reduce the ratio of their investments into TSE
 whenever the gold price is decreasing.

{\bf Barter when facing sanctions!} 
Our finding has also the important implications for policy maker of 
countries faced international sanctions. 
The lack of dependence between the TSE index and the oil price 
generating the major foreign currency revenue 
suggests that oil dollars almost do not influence the national market
of Iran.   
 
Faced with the international sanctions, 
Iran was turning to barter by offering gold bullion
 in overseas vaults or tankerloads of oil, 
 in return for food as the financial sanctions 
 had hurt its ability to import basic staples \cite{Reuters:2012}.
 Those sanctions were not banning companies from selling food to Iran, 
 but the transactions with banks were very difficult.
 Unable to bring in U.S. dollars and euros ahead of the new U.S. sanctions,
  Iran is open to accepting agricultural products and medical equipment
   in exchange for its crude oil \cite{oilprice:2018}. 
A scheme to barter Iranian oil for European goods through Russia, 
which would then refine it and sell it to Europe
as part of mechanism to bypass American sanctions on the Islamic Republic
was announced by Europe, China, and Russia along the sidelines of the United Nations General Assembly \cite{independent:2018}.

In the future, we plan to implement the copula method by using rolling windows for modeling the time-varying dependence between gold, oil and TSE in a multivariate framework.


\begin{thebibliography}{999}
	
	
	\bibitem{Guzman:2013}
	Timothy Alexander Guzman (10 April 2013). "New Economic Sanctions on Iran, Washington's Regime Change Strategy". Global Research. Retrieved 5 May 2013.
	
	\bibitem{Blanchard:2007}
	Blanchard, Olivier, and Jordi Gali.  2007.  "The Macroeconomic Effects of Oil Shocks: Why Are the 2000s So Different from 1970s."  NBER Working Paper No. 13368.
	
	
	\bibitem{FT:2019}
	Financial Tribune,
	Iran's Proven Oil Reserves to Rise by 10 Percent, Thursday March 09 2019.
	Available at 
{https://financialtribune.com/articles/energy/82929/irans-proven-oil-reserves-to-rise-by-10-percent} 
	
	\bibitem{Press.TV:2010}
	Press.TV Iran oil exports top 844mn barrels
	Wed Jun 16, 2010 4:59PM. 
	Available at 
	 {https://web.archive.org/web/20120311194302/http://edition.presstv.ir/detail/130736.html}
	
	\bibitem{Nemat:2016}
	Nemat Falihy Pirbasti,   Mehdi Tajeddini, 
	"Factors Affecting on the Price of Gold on Global Markets and Its
	Impact on the Price of Gold in Iran Market (Incorporation of Dynamic
	System Pattern and Econometric)",
	{\it Modern Applied Science} Vol. {\bf 10}, No. 3; 
	Published by Canadian Center of Science and Education
	(2016).
	
	\bibitem{FT:2018}
	Financial Tribune,
	Economy, Business And Markets
	August 04, 2018 19:51
	"Gold Demand at Four-Year High". 
	Available at 
{https://financialtribune.com/articles/economy-business-and-markets/91119/gold-demand-at-four-year-high}
	
	
	\bibitem{LFP}
	Historical London Fix Prices.
	Available at 
{https://www.kitco.com/gold.londonfix.html}
	
	 \bibitem{TSE}
	The current information about Tehran Stock Exchange is available at {www.tse.ir}
	
	
	
	\bibitem{Hamilton:1983}
	Hamilton, J. D. (1983). Oil and the macroeconomy since World War II. \textit {Journal of Political Economy}.
	\textbf{91} (2): pp. 228-48.
	
	
	\bibitem {Kling:1985}
	Kling, J. L. (1985). Oil price shocks and stock market behavior. \textit {Journal of Portfolio Management}. \textbf 12:
	pp. 34-39.
	
	\bibitem {Huang:1996}
	Huang, R., Masulis, R., Stoll, H. (1996). Energy shocks and financial markets. \textit {Journal of Futures
	Markets}. \textbf{16} (1): pp. 1-27.
	
	\bibitem{Sadorsky:1999}
	Sadorsky, P. (1999). Oil price shocks and stock market activity, \textit {Energy Economics}, Vol. \textbf {21},
	pp.449- 469.
	
	\bibitem{Cai:2011}
	Cai, J., Y.L., Cheung, M.C., Wong. (2011). What moves the gold market?, \textit {Journal of Future Markets}, \textbf{21}: pp. 257 - 278
	
	\bibitem{Baryshevsky:2004}
	Baryshevsky, D. V. (2004). "The Interrelation of the Long-Term Gold Yield with the Yields of Another Asset Classes." from http://ssrn.com/abstract=652441.
	
	\bibitem{Nandha:2008}
	Nandha, M., and Faff, R. (2008). Does oil move equity prices? A global view. \textit{Energy Economics}. \textbf {30}, 986-997.
	
	\bibitem{Miller:2009}
	Miller, J. I., Ratti, R. A. (2009). Crude oil and stock markets: Stability, instability, and bubbles. \textit {Energy Economics}, \textbf {31}(4), 559-568.
	
	\bibitem{Baur:2010}
	Baur, D.G. and Mcdermott, T.K. (2010). Is gold a safe haven? International evidence.  \textit {Journal of Banking \& Finance}, \textbf {34} (8), 1886-1898.
	
	\bibitem{Batten:2010}
	Batten, J. A., Ciner, C., Lucey, B. M. (2010). The Macroeconomic Determinants of Volatility in Precious Metals Markets. \textit {Resources Policy}, vol. \textbf{35}(2), 65-71.
	
	\bibitem{Bharn:2010}
	Bharn, R., Nikolovann, B. (2010). Global oil prices, oil industry and equity returns: Russian experience. \textit {Scottish Journal of Political Economy}, \textbf {57}(2), 169-186.
	\bibitem{Filis:2011}
	Filis, G., Degiannakis, S., \& Floros, C. (2011). Dynamic correlation between stock market and oil prices:
	The case of oil-importing and oil-exporting countries. \textit {International Review of Financial Analysis}, \textbf{20}(3),
	152 - 164.
	
	\bibitem{Arouri:2011b}
	Arouri, M.E.H., Lahiani, A. and Nguyen, D.K. (2011b). Return and volatility transmission between world oil prices and stock markets of the GCC countries, \textit {Economic Model}, Vol. \textbf {28}, pp.1815 - 1825.
	
	\bibitem{Chang:2012}	
	Chang, K.L. (2012). The Time-Varying and Asymmetric Dependence between Crude Oil Spot and Futures Markets: Evidence from the Mixture Copula-based ARJI - GARCH Model, \textit{Economic Modelling}, \textbf{29}(6), 2298 - 2309.
	
	\bibitem{Jaschke:2012}
	S. J\"{a}schke, K.F. Siburg, P.A. Stoimenov, (2012), Modeling dependence of extreme events in energy markets using tail copulas, \textit{J. Energy Markets}, \textbf{5}(4): 63-80.
	
		
	\bibitem{Nguyen:2012}
	Nguyen, C. and Bhatti, M.I. (2012). Copula model dependency between oil prices and stock markets: evidence from China and Vietnam, \textit{Journal of International Financial Markets. Institutions and Money}, Vol. \textbf{22}, pp.758 - 773.
	
	\bibitem{Aloui:2013}
	Aloui, R., Hammoudeh, S., Nguyen, D. K. (2013). A time-varying copula to oil and stock market dependence: The case of transition economies. \text {Energy Economics}, \textbf {39}, 208 - 221.
	
	\bibitem{Mensi:2013}
	Mensi, W., Beljid, M., Boubaker, A., Managi, S. (2013). Correlations and volatility spillovers across commodity and stock markets: linking energies, food, and gold. \textit {Econ. Modell}. \textbf{32}, 15 - 22.
		
	\bibitem{Silva:2014}
	Silva, P. P., Rebelo, P., Afonso, C., 2014. Tail dependence of financial stocks and CDS markets:
	Evidence using copula methods and simulation-based inference. \textit {Economics-The Open-Access,
	Open-Assessment E-Journal}. \textbf{8}(39): 1-27.
	
	\bibitem{Arouri:2014}
	Arouri, M.E.H., Lahiani, A. and Nguyen, D.K. (2014). World gold prices and stock returns in China: insights for hedging and diversification strategies, \textit {Econ. Modell.}, \textbf{44}, pp. 273-282
	
	\bibitem{Zhu:2016b}
	Zhu, H., Huang, H., Peng, C., Yang, Y. (2016b). Extreme dependence between crude oil and stock markets in Asia-Pacific regions: Evidence from quantile regression. \textit {Economics Discussion Papers}, No \textbf{2016-46},
	Kiel Institute for the World Economy. http://www.economics-ejournal.org/economics/discussionpapers/2016-46
	
	\bibitem{sima:2017}
	Siami-Namini, S.,  Hudson, D. (2017). Volatility Spillover Between Oil Prices, Us Dollar Exchange Rates And International Agricultural Commodities Prices. presented at \textit {2017 Annual Meeting}, February 2017, Mobile, Alabama 252845, Southern Agricultural Economics Association.
	
	
	\bibitem{Trabelsi:2016b}
	Trabelsi, N. (2017). Asymmetric tail dependence between oil price shocks and sectors of Saudi Arabia System. \textit{The Journal of Economic Asymmetries}, \textbf{16} 26 - 41
	
	\bibitem{Hamma:2018}
	Hamma,W. , Ghorbel , A., Jarboui, A., (2018). Copula model dependency between oil prices and stock markets: evidence from Tunisia and Egypt. \textit{American Journal of Finance and Accounting}. \textbf{2} pp. 111-150
	
	\bibitem{Foster:2008}
	Foster, K.R, Kharazi, A. (2008). Contrarian and momentum returns on Iran's Tehran Stock Exchange. \textit Journal of International Financial Markets Institutions and Money \textbf{18}(1):16-30
	
	\bibitem {Najafabadi:2012}
	Najafabadi, A.T.P., Qazvini, M., Ofoghi, R. (2012) The impact of oil and gold prices shock on Tehran stock exchange: a copula approach. Iran. \textit{J. Econ. Stud}. \textbf {1}(2), 23 - 47 
	
	\bibitem{Shams:2014}
	Shams, S., Zarshenas, M. (2014). Copula approach for modeling oil and gold prices and exchange rate co-movements in Iran \textit {Int. J. Stat. Appl.}, \textbf{4}(3), pp. 172-175
	
	\bibitem{Fuller:1976}
	Fuller,W.A. (1976). \textit{Introduction to Statistical Time Series}. New York: John Wiley and Sons.
	
	\bibitem{Hamilton:1994}
	Hamilton, J. D. (1994). \textit{Time series analysis}. Princeton, N.J.: Princeton University Press. 
	
	\bibitem{Sklar:1959}
	Sklar, A. (1959), Fonctions de r\'{e}partition \'{a} $n$-dimensions et leurs marges, \textit {Publ. Inst. Statist. Univ. Paris} (in French), \textbf{8}: 229 - 231.
		
	\bibitem{Nelsen:1999}
	Nelsen, R. B. (1999). \textit{An Introduction to Copulas}. Springer, New York.
	
	\bibitem{Embrechts:1999}
	Embrechts, P., McNeil, A. J., and Straumann, D. (1999). Correlation: pitfalls and alternatives. \textit {Risk Magazine}, \textbf {5}:69-71.
	
	\bibitem{Schmidt:2006}
	Schmidt, R., Stadtm\"{u}ller, U. (2006). Nonparametric estimation of tail dependence. \textit {Scandinavian Journal of Statistics}, \textbf{32}(2), 307 - 335.
	\bibitem{Ljung:1978}
	Ljung, G. M., Box, G. E. P.  (1978). On a Measure of a Lack of Fit in Time Series Models. \textit{Biometrika}. \textbf{65} (2): 297 - 303
			
	\bibitem{Ghalanos:2018}
	Alexios Ghalanos (2018). \textit{rugarch: Univariate GARCH models}. R package version 1.4--0.
	
	\bibitem{Diebold:1998}
	Diebold FX, Gunther TA, Tay AS. (1998). Evaluating density forecasts, with applications to financial risk management. \textit {International Economic Review} \textbf{39}: 863 - 883.
	
	
	\bibitem{Kolmogorov:1933}
	Kolmogorov A. (1933): Sulla determinazione empirica di una legge di distribuzione. —\textit{ G. Ist. Ital.
	Attuari}, vol.\textbf{4}, pp. 83–91
	
	\bibitem{Kosik:1985}
	Kosik P., Sarkadi K. (1985). A new goodness-of-fit test \textit{ Proc. of 5-th Pannonian Symp. of Math. Stat.},
	Visegrad, Hungary, \textbf{20}: P. 267 272
	
	\bibitem{uniftest:2015}
	Melnik, M., Pusev, R. (2015). \textit{uniftest: Goodness-of-fit tests for the uniform distribution}. R package version 1.1.
	
	
	\bibitem{Deheuvels:1979}
	Deheuvels, P. (1979). La fonction de d\'{e}pendance empirique et ses propri\'{e}t\'{e}s. Un test non param\'{e}trique d'ind\'{e}pendance. \textit{Acad. Roy. Belg. Bull. Cl. Sci.} \textbf{65}(6), 274 - 292.	
	
	\bibitem{Genest:2007}
	Genest, C., Favre, A.-C., 2007. Everything you always wanted to know about copula modeling but were afraid to ask. \textit {Journal of Hydrologic Engineering}, \textbf{12}, 347 - 368.
	
	
	\bibitem{Genest:2011}
	Genest, C., Kojadinovic, I., Neslehov\'{a},J., Yan, J. (2011). A goodness-of-fit test for bivariate extreme-value copulas. \textit {Bernoulli}, \textbf{17}(1): pp. 253-275
	
	\bibitem{vinecopula:2011}
	 Schepsmeier U., Stoeber, J., Brechmann, E.C., Graeler, B., 
	 {\it Package VineCopula: Statistical Inference of Vine Copulas}
	 R-Project CRAN Repository (2013).
	 
	
	
	\bibitem{Genest:2009}
	Genest, C., R\'{e}millard, B.  Beaudoin, D.  (2009). Goodness-of-fit tests for copulas: A review and a power study. \textit {Insurance Math. Econom}. \textbf {44}(2): 199 - 213.
	
	\bibitem{Genest:2008}
	Genest, C. and R\'{e}millard, B.  (2008). Validity of the parametric bootstrap for goodness-of-fit testing in semiparametric models. \textit {Annales de l'Institut Henri Poincar\'{e} - Probabilit\'{e}s et Statistique} \textbf {44}(6): pp. 1096 - 1127.
	
		
\bibitem{Reuters:2012}
Reuters World News,
February 9, 2012 / 5:59 AM;  
"Iran turns to barter for food as sanctions cripple imports"
by Valerie Parent, Parisa Hafezi.
Available at 
{https://www.reuters.com/article/us-iran-wheat/iran-turns-to-barter-for-food-as-sanctions-cripple-imports-idUSTRE8180SF20120209}

	\bibitem{oilprice:2018}
	oilprice.com, 
	"Iran Looks To Barter Oil As U.S. Sanctions Bite" by
	Tsvetana Paraskova - Jul 05, 2018, 6:00 PM CDT. Available at 
	{https://oilprice.com/Geopolitics/International/Iran-Looks-To-Barter-Oil-As-US-Sanctions-Bite.html}
	
\bibitem{independent:2018}
independent.co.uk, 
"Europe and Iran plot oil-for-goods scheme to bypass US sanctions"
by Borzou Daragahi, 
Wednesday 26 September 2018 21:58; 
Available at 
{https://www.independent.co.uk/news/world/middle-east/iran-sanctions-trump-nuclear-deal-europe-russia-oil-un-a8556786.html}


	
\end{thebibliography}
\end{document}